\documentclass[]{spie}  

\usepackage{amsmath}
\usepackage{amssymb}
\usepackage[ansinew]{inputenc}
\usepackage{enumerate} 
\usepackage{url}
\usepackage{multicol}

\newcommand{\mb}[1]{\mathbb{#1}}

\newcommand{\mf}[1]{\mathfrak{#1}}
\newcommand{\mr}[1]{\mathrm{#1}}

\newcommand{\braket}[2]{\left \langle #1 \, \left | \right . \, #2 \right \rangle }

\newcommand{\ket}[1]{\left | \, #1 \right \rangle}

\newcommand{\modulo}[1]{\left | #1 \right |}
\newcommand{\modulop}[2]{\left | #1 \right |^{#2}}

\newcommand{\urlbib}[1]{{\footnotesize{\url{#1}}}}

\title{A possible hypercomputational quantum algorithm}

\author{Andrés Sicard, Mario Vélez, and Juan Ospina
\skiplinehalf
Logic and Computation Group, EAFIT University,  A.A. 3300, Medellín, Colombia
}

\authorinfo{Further author information: (Send correspondence to A.S.)
\\
A.S.: E-mail: asicard@eafit.edu.co, Telephone: 57 4 261 95 00
\\  
M.V: E-mail: mvelez@eafit.edu.co, Telephone: 57 4 261 95 00
\\
J.O: E-mail: judoan@epm.net.co, Telephone: 57 4 261 95 00}

\begin{document} 
  \maketitle 

\begin{abstract}
The term `hypermachine'  denotes any data processing device (theoretical or that can be implemented) capable of carrying out tasks that cannot be performed by a Turing machine. We present a possible quantum algorithm for a classically non-computable decision problem, Hilbert's tenth problem; more specifically, we present a possible hypercomputation model based on quantum computation. Our algorithm is inspired by the one proposed by Tien D. Kieu, but we have selected the infinite square well instead of the (one-dimensional) simple harmonic oscillator as the underlying physical system. Our model exploits the quantum adiabatic process and the characteristics of the representation of the dynamical Lie algebra $\mf{su}(1,1)$ associated to the infinite square well. 
\end{abstract}




\section{INTRODUCTION}
\
\begin{flushright}
\emph{Once upon on time, back in the golden age of 
\\
the recursive function theory, computability was an absolute.}
\\
Richard Sylvan and Jack Copeland\cite{Sylvan-Copeland-2000}.
\end{flushright}
\
The hypercomputers, according to Copeland and Proudfoot \cite{Copeland-Proudfoot-1999a}, compute functions or numbers, or more generally, solve problems or carry out tasks, that cannot be computed or solved by a Turing machine (TM). The hypercomputation theory rejects the idea of an \emph{absolute computability} (i.e. Turing computability), detached from logical, mathematical, physical or biological theories. Notwithstanding, the up coming of an academic community that works the concept of hypercomputation\footnote{Conferences: \emph{The Hypercomputation Workshop} in London, May 24, 2000 (\url{www.alanturing.net/turing_archive/conference/hyper/hypercom.html}); \emph{Beyond the Classical Boundaries of Computation}, session of the American Mathematical Society Meeting in San Francisco, May, 2003 (\url{www.ams.org/notices/200305/sanfran-prog.pdf}). Published special issues: Vols. 12(4) and 13(1) of \emph{Mind and Machines}; Vol. 317 of \emph{Theoretical Computer Science}. Web pages: \url{www.hypercomputation.net}.}, and notwithstanding the proliferation of theoretical hypercomputation models, \cite{Copeland-2002} the possibility of \emph{real} construction of a hypermachine is controversial and is still under analysis.

At first glance, we could think that the possibility of a hypercomputation model would be a refutation of the widely accepted Church-Turing thesis, which is usually interpreted as the identification of the \emph{naturally} calculable functions with the TM-computable functions; but actually the existence of hypercomputation models refute the thesis M, which identifies the functions \emph{calculable by a machine} with the TM-computable functions \cite{Sicard-Velez-2001a-english}. In other words, the existence of the current proposed hypercomputation models coexists with acceptance of the Church-Turing thesis.

The existence of supported proposals of hypercomputation in quantum mechanics is ample \cite{Copeland-2002}, nevertheless this is not the case for those based on quantum computation. Considering the fact that quantum computation extends beyond its ``standard''
model (i.e. quantum Turing machines or quantum circuits) and it includes proposals such as continuous \cite{Lloyd-Braunstein-1999}, adiabatic \cite{Farhi-Goldstone-Gutman-Lapan-Lundgren-Preda-2001}, and/or holonomic quantum computation \cite{Niskanen-Nakahara-Salomaa-2003}, among others; this article presents the construction of a hypercomputation model from quantum computation. Our model is based on the one proposed by Tien D. Kieu \cite{Kieu-2003c,Kieu-2003a,Kieu-2003b}, but we have selected the infinite square well (ISW) instead of the (one-dimensional) simple harmonic oscillator (SHO) as
the underlying physical system. Our model exploits the quantum adiabatic process, and due to this change, our model is supported by some characteristics of the dynamical Lie algebra $\mf{su}(1,1)$, associated to the ISW; instead of the dynamical Lie algebra Weyl-Heisenberg $\mf{g}_{\mr{W \negthickspace - \negthickspace H}}$, associated to the SHO. 

\section{Kieu's algorithm}
A Diophantine equation is of the following form
\begin{equation}\label{eq-1}
D(x_1, \dots, x_k) = 0  ,   
\end{equation}
where $D$ is a polynomial with integer coefficients. In present terminology, Hilbert's tenth problem may be paraphrased as: Given a Diophantine equation of type \eqref{eq-1}, we should build a procedure to determine whether or not this equation has a solution in non-negative integers $\mb{N}$. From the concluding results obtained by Matiyasevich, Davis, Robinson, and Putnam, we know that, in the general case, this problem is algorithmically insolvable or more precisely, it is TM incomputable \cite{Matiyasevich-1993}. 

The hypercomputability of Kieu's algorithm is due to the fact that this algorithm solves Hilbert's tenth problem. From its preliminary version \cite{Kieu-2001}, Kieu has presented different refinements to his
algorithm \cite{Kieu-2003c,Kieu-2003a,Kieu-2003b,Kieu-2002b,Kieu-2003,Kieu-2004a,Kieu-2005a}\footnote{In the Discussion List FOM - Foundations of Mathematics (\url{www.cs.nyu.edu/pipermail/fom/}) some characteristics of this algorithm have been analyzed and discussed.}. In addition, with basis in his algorithm, Kieu has reformulated Hilbert's tenth problem in strictly mathematical terms, indicating a
possible way of solution through the mathematical analysis and the theory of infinite-dimensional operators \cite{Kieu-2004}. 

Kieu's algorithm is constructed beginning with the SHO characteristics. For the SHO with Hamiltonian 
\begin{align}\label{eq-3}
  \begin{split}
    H &= (P^2 + X^2)/2
      \\
      &= a^{\dagger}a + 1/2 \enspace,
\end{split}
\end{align}
the occupation-number states $\ket{n}$,  the action of the creation  $a^\dagger$ and annihilation $a$ operators on occupation-number states, their commutation relations $[a,a^\dagger], [a,a]$, and $[a^\dagger, a^\dagger]$, the occupation-number operator $N$ and the coherent states $\ket{\alpha}$ are given by
\begin{gather}
\{\ket{n} \mid n \in \mb{N} \} \enspace, \label{eq-5}  
\\
a \ket{0} = 0 \enspace,  \qquad a \ket{n} = \sqrt{n}\ket{n-1} \enspace, \qquad a^\dagger \ket{n} = \sqrt{n+1}\ket{n+1} \enspace,\label{eq-10} 
\\
[a,a^\dagger] = 1 \enspace, \qquad [a,a] = [a^\dagger, a^\dagger] = 0 \enspace, \label{eq-25}  
\\
N = a^\dagger a \enspace, \label{eq-30} 
\\
 \ket{\alpha} =  e^{-\frac{|\alpha|^2}{2}}\sum_{n=0}^{\infty}
\frac{\alpha^n}{\sqrt{n!}}\ket{n} \enspace, \qquad \alpha \in \mb{C} \enspace, \label{eq-35}
\end{gather}
With the basis for the SHO, Kieu's algorithm \cite{Kieu-2003c} is shown in Table \eqref{tabla-10}.

\begin{table}[ht]
\caption{\footnotesize{Kieu's hypercomputational quantum algorithm.}}
\label{tabla-10}
\begin{center}
\begin{tabular}{|p{16cm}|}
\hline
Given a Diophantine equation with $k$ unknowns of type \eqref{eq-1}, Kieu provides the following quantum algorithm to decide whether this equation has any non-negative integer solution or not:
\begin{enumerate}[1.]
\item Construct a physical process in which a system initially
starts with a direct product of $k$ coherent states $\ket{\psi(0)}$, and in which the system is subject to a time-dependent Hamiltonian $H_{\mr{A}}(t)$ over the
time interval $[0,T]$, for some time $T$, with the initial Hamiltonian $H_{\mr{I}}$ and the final Hamiltonian $H_{\mr{D}}$, given by
{\begin{align}
\ket{\psi(0)} &= \bigotimes_{i=1}^k \ket{\alpha_i}  \enspace, \label{eq-60}
\\
H_{\mr{A}}(t) &= (1 - t/T)H_{\mr{I}} +  ( t/T) H_{\mr{D}} \enspace, \label{eq-70}
\\
H_{\mr{I}} &= \sum_{i=1}^k \left(a_i^\dagger - \alpha^*_i\right)
\left(a_i - \alpha_i\right) \enspace, \label{eq-80}
\\
H_{\mr{D}} &= \left(D(N_1, \ldots, N_k)\right)^2 \enspace. \label{eq-90}
\end{align}}
\item Measure through the time-dependent Schr\"odinger equation 
{\begin{equation*}
  i\partial_t \ket{\psi(t)} = H_{\mr{A}}(t) \ket{\psi(t)}, \mbox{ for $t \in [0,T]$}
\end{equation*}}
the maximum probability to find the system in a
particular occupation-number state at the chosen time $T$,
{\begin{align*}
    P_{\mr{max}}(T) &= \max_{(n_1,\dots,n_k) \in \mb{N}^k} \left | \braket{\psi(T)}{n_1,\dots,n_k}  \right |^2 ,
\\
&=\left | \braket{\psi(T)}{\{n\}}_0 \right | ^2 ,
\end{align*}}
where $\ket{\{n\}}_0$ (which is a direct product of $k$ particular occupation-number state, $\bigotimes_{i=1}^k \ket{n_i}_0$) provides that maximal probability among all other direct products of $k$ occupation-number states.

\item If $P_{\mr{max}}(T) \le 1/2$, increase $T$ and repeat all the steps above.
\item If 
  \begin{equation}\label{eq-93}
    P_{\mr{max}}(T)>1/2     
  \end{equation}
then $\ket{\{n\}}_0$ is the ground state of $H_{\mr{D}}$ 
(assuming
no degeneracy) and we can terminate the
algorithm and deduce a conclusion from the fact that
\begin{center}
\fbox{\parbox[t]{11cm}{
$H_{\mr{D}}\ket{\{n\}}_0 = 0$ iff the equation \eqref{eq-1} has a 
non-negative integer solution.
}}
\end{center}
\end{enumerate}
\\
\hline
\end{tabular}
\end{center}
\end{table}

Kieu has indicated the following characteristics of his algorithm \cite{Kieu-2003c,Kieu-2003a,Kieu-2003b}:
\begin{enumerate}

\item The algorithm is probabilistic such as are quantum algorithms in general.

\item The infinite-dimension Hamiltonian \eqref{eq-3} acts on some Fock space whose orthonormal basis is \eqref{eq-5}, and satisfies that $\ket{n} = \frac{ (a^{\dagger})^n}{\sqrt{n!}} \ket{0}$.

\item The codification of \eqref{eq-1} is made by the occupation-number operator \eqref{eq-30} which is diagonal on the basis \eqref{eq-5}, and whose eigenvalues are the non-negative integers.

\item The application of the adiabatic theorem to obtain from the ground state \eqref{eq-60} associated with the eigenvalue zero of \eqref{eq-80}, the ground state $\ket{\{n\}}_0$ of \eqref{eq-90}. 

\item  The finiteness of the run time $T$ to obtain with a high probability the
ground state $\ket{\{n\}(T)}_0$ of $H_D$. Although the value of $T$ is not calculable a priori, its finiteness is guaranteed because the ground state of $H_A(sT)$ for $0 \leq s < 1$ is non-degenerate, and this ground state never crosses with any other state during the adiabatic regime. 

\item The halting criterion \eqref{eq-93} is established by the maximum
peak of density of probability associated with the initial state \eqref{eq-60}, where for any $n$ and $\alpha$ 
\begin{align*}
  \modulop{\braket{\alpha}{n}}{2} &= e^{|\alpha|^2}\frac{|\alpha|^{2n}}{n!} 
  \\
  & < 1/2 \enspace.
      \end{align*}
In addition, it is necessary to verify that the probability of any excited state cannot be greater than $1/2$ at anytime.


\end{enumerate}

\section{Hypercomputational quantum algorithm \emph{à la} Kieu}
Although the physical referent used by Kieu for the construction of his algorithm was the SHO, parting from the description on Table \eqref{tabla-10}, it can be observed that its constitutive elements, come forth from the dynamical algebra Weyl-Heisenberg $\mf{g}_{\mr{W \negthickspace - \negthickspace H}}$ associated with the SHO. The dynamical algebra $\mf{g}_{\mr{W \negthickspace - \negthickspace H}}$ satisfies the commutation relations \eqref{eq-25}. Beginning with its generating elements $a, a^{\dagger}$ and $1$, it is possible to factor the Hamiltonian from the SHO as pointed out by \eqref{eq-3} and it is possible to construct the occupation-number operator \eqref{eq-30}. The coherent states \eqref{eq-35} correspond to the eigenstates of annihilation operator $a$, in other words, $a\ket{\alpha} = \alpha \ket{\alpha}$. Furthermore, the algebra $\mf{g}_{\mr{W \negthickspace - \negthickspace H}}$ has an infinite-dimensional irreducible representation wherein the action of its generators on basis \eqref{eq-5} is given by \eqref{eq-10}. This representation makes posible that the eigenvalues of operator \eqref{eq-30} are the non-negative integers and its eigenstates are the basis \eqref{eq-5}, in other words $N\ket{n} = n\ket{n}$. 

To carry out the construction of our algorithm \emph{à la} Kieu, we have selected as physical referent the ISW, which has an associated dynamical algebra different to the one used by Kieu. For a particle with mass $m$ trapped inside the infinite well $0 \leq x \leq \pi l$, the Hamiltonian operator $H^{\mr{ISW}}$ and the energy levels  $E_{n}^{\mr{ISW}}$ are \cite{Antoine-Gazeaub-Monceauc-Klauder-Penson-2001}
\begin{equation}\label{eq-100}
   H^{\mr{ISW}} = i^2 \frac{\hbar^2}{2m}\frac{d^2}{dx^2} - \frac{\hbar^2}{2ml^2} \enspace, \qquad E_n^{\mr{ISW}} = \frac{\hbar^2}{2ml^{2}}n(n+2) \enspace , \end{equation}
where the action of $H^{\mr{ISW}}$ on basis \eqref{eq-5} is
\begin{equation*}
  H^{\mr{ISW}} \ket{n} = E_n^{\mr{ISW}} \ket{n} \enspace .
\end{equation*}

Due to the spectral structure of the ISW, the dynamical algebra associated with it, is the Lie algebra $\mf{su}(1,1)$ \cite{Antoine-Gazeaub-Monceauc-Klauder-Penson-2001}. This is a three-dimensional algebra that satisfies the commutation relations
\begin{equation*}
 [K_-,K_+] = K_3 \enspace, \qquad [K_-,K_3] = 2K_- \enspace,  \qquad [K_+,K_3] = -2K_+ \enspace ,
\end{equation*} 
where operators $K_+, K_-$ and $K_3$ are called creation, annihilation and Cartan operators, respectively. The algebra $\mf{su}(1,1)$ admits an infinite-dimensional irreducible representation where actions of $K_+, K_-$ and $K_3$ on basis \eqref{eq-5} are
\begin{equation}\label{eq-106}
  \begin{aligned}
    K_+\ket{n} &= \sqrt{(n+1)(n+3)}\ket{n+1} \enspace , & K_-\ket{0} &= 0 \enspace , & K_3\ket{n} &= (2n+3)\ket{n} \enspace . 
\\
& & K_-\ket{n} & = \sqrt{n(n+2)}\ket{n-1} \enspace ,
\end{aligned}
\end{equation}

With basis in the algebra $\mf{su}(1,1)$, the Hamiltonian \eqref{eq-100} is rewritten as 
\begin{equation}\label{eq-110}  
H^{\mr{ISW}} = \frac{\hbar^2}{2ml^{2}}K_+K_- \enspace, 
\end{equation}
and a new number operator $N^{\mr{ISW}}$ is given by
\begin{equation}\label{eq-118} 
N^{\mr{ISW}} = (1/2)(K_3 - 3) \enspace,  
\end{equation}
where $N^{\mr{ISW}}\ket{n} = n\ket{n}$, in other words, the operator $N^{\mr{ISW}}$ presents the same spectral characteristics as the occupation-number operator \eqref{eq-30}.

Due to the dynamical algebra associated, the Barut-Girardello coherent states $\ket{z}, z \in \mb{C}$, for the ISW are eigenstates of annihilation operator $K_-$ \cite{Wang-Sanders-Pan-2000}
\begin{equation}\label{eq-120}
\ket{z} = \frac{\modulo{z}}{\sqrt{I_2(2\modulo{z})}}\sum_{n=0}^{\infty}\frac{z^n}{\sqrt{n!(n+2)!}}\ket{n} \enspace,
\end{equation}
where $I_v(x)$ is the modified Bessel function of the first kind.

With the presented elements, Kieu's algorithm  is
rewritten in the following way. Instead of replacing each one of the variables of \eqref{eq-1} by \eqref{eq-30} to construct \eqref{eq-90}, these can be
replaced by \eqref{eq-118} to obtain
\begin{equation}\label{eq-130}
 H_{D}^{\mr{ISW}}=\left ( D \left( N_1^{\mr{ISW}}, \dots, N_k^{\mr{ISW}} \right ) \right ) ^{2} \enspace .
\end{equation}

Due to \eqref{eq-130}, it is necessary to construct a new initial Hamiltonian $H_I^{\mr{ISW}}$ from  creation and annihilation \eqref{eq-106} operators of $\mf{su}(1,1)$
\begin{equation}\label{eq-135}
H_I^{\mr{ISW}} = \sum_{i=1}^k (K_{+_i} - z_i^*)(K_{-_i} - z_i) \enspace ,
\end{equation}
which has associated with the eigenvalue zero, the ground state $\ket{\psi^{\mr{ISW}}(0)}$, constructed from the coherent states \eqref{eq-120}
\begin{equation}\label{eq-140}
\ket{\psi^{\mr{ISW}}(0)} = \bigotimes_{i=1}^k \ket{z_i} \enspace .
\end{equation}

Finally, from \eqref{eq-130} and \eqref{eq-135} the Hamiltonian \eqref{eq-70} takes the form
\begin{equation}\label{eq-150}
H_A^{\mr{ISW}}(t) = \left(1 - \frac{t}{T} \right)H_I^{\mr{ISW}} + \left( \frac{t}{T} \right ) H_D^{\mr{ISW}} \enspace.
\end{equation}

Some observations with respect to the new algorithm: 
\begin{enumerate}

\item The infinite-dimension Hamiltonian \eqref{eq-110} acts on some Fock space too, whose orthonormal basis is \eqref{eq-5}, and satisfies that $\ket{n} = \frac{(K_+)^n}{\sqrt{n!(n+2)!/2}} \ket{0}$.

\item The Hamiltonians \eqref{eq-70} and \eqref{eq-150} are unbounded operators, therefore it is necessary to use a version of the adiabatic theorem for unbounded operators \cite{Avron-Elgart-1999}, for both algorithms. However, this fact is not very significant because the algorithm operates more in the infrared part of the spectrum  than in the ultraviolet part. 

\item The conditions required for the finiteness of the run time $T$ are satisfied by \eqref{eq-150} adapting Kieu's arguments for algebra $\mf{g}_{\mr{W \negthickspace - \negthickspace H}}$ \cite{Kieu-2003c} to algebra $\mf{su}(1,1)$. 

\item In order to satisfy the halting criterion \eqref{eq-93}, it is necessary to select $z_i$ values such that $|z_i| > 1.6$ for the construction of \eqref{eq-140}, since according to the density of probability associated to the
states  \eqref{eq-120} for any $n$ and $|z| > 1.6$
\begin{align*}
\modulop{\braket{z}{n}}{2} &= \frac{|z|^2}{2I_2(2\modulo{z})} 
\\
& < 1/2 \enspace,
\end{align*}
In the simulations of our algorithm \cite{Sicard-Ospina-Velez-2005}, this argument has been numerically confirmed. 


\item Unlike the algebra $\mf{g}_{\mr{W \negthickspace - \negthickspace H}}$, the codification of \eqref{eq-1} in the algebra $\mf{su}(1,1)$ could be directly carried out with the diagonal opeator  $K_3$. Apparently, due to  \eqref{eq-106} this codification change, changes the problem to be resolved, given that instead of establishing if  \eqref{eq-1} has, or doesn't have non-negative integers solutions, the only thing that could be established is if \eqref{eq-1} has or doesn't have solutions of the form $2n+3$, with $n \in \mb{N}$. Nevertheless, this problem is equivalent to Hilbert's tenth problem parting from the construction of an finite system of Diophantine equations \cite{Matiyasevich-1993}. 

\item Unlike the algebra $\mf{g}_{\mr{W \negthickspace - \negthickspace H}}$, the algebra $\mf{su}(1,1)$ admits different types of coherent states (Barut-Girardello, Klauder, Perelomov, etc.) \cite{Wang-Sanders-Pan-2000}. The advantage of a type of coherent state over another one would be in computational complexity issues, rather than in computational power issues.

\item  The recent equivalence between the adiabatic computation and the ``standard'' quantum computation  \cite{Aharonov-van--Dam-Kempe-Landau-Lloyd-Regev-2004} does not generate any contradiction with the hypercomputational characteristics of our algorithm (or Kieu's algorithm), given that our algorithm is one of adiabatic quantum computation over infinite-dimensional spaces, and the demonstration of equivalence indicated is for the finite-dimensional case.

\item The Lie algebra $\mf{su}(1,1)$ is the dynamical algebra associated with different physical referents such as the infinite cylindrical wells, the Pöschl-Teller potentials, quantum optics systems with $SU(1,1)$ symmetries, among others. Therefore, in principle, it is possible to select one of these referents as an underlying physical system of our hypercomputational quantum algorithm \cite{Sicard-Ospina-Velez-2005a}. 

\end{enumerate}

\section{NON-(Turing machine) hypercomputation}
Within the context of hypercomputation, a hypermachine also capable of simulating a universal Turing machine is called a super-TM, otherwise it is called a non-TM \cite{Stannett-2003}. The possible universality of our hypercomputation model based on the ISW would be established by its capability to generate a set of quantum gates, such that any unitary transformation $U(2^n)$, that is, any quantum gate that operates upon $n$-qubits, can be approximated with sufficient exactness by a quantum circuit that is only made of a finite number of gates of this set.

For example\footnote{Henceforth, we will use the convention of  a superindex over the operators (quantum gates) and over the states (qubits). This superindex will denote the dimension of Hilbert space upon which the operators act or upon which the states are defined.}, for the controlled-NOT gate ($CNOT^4$), the transformation carried out upon the $2$-qubit canonical basis $\{\ket{00}^4,\ket{01}^4, \ket{10}^4, \ket{11}^4\}$, is given by
 \begin{equation}\label{eq-universalidad-01}
 \ket{x,y}^4   \xrightarrow{CNOT^4}  \ket{x,x \oplus y}^4\enspace.
\end{equation}

For the ISW with the Hamiltonian independent from the time \eqref{eq-100}, the states of the system evolve according to the Schrödinger equation solution of stationary states \cite{Messiah-1990}
\begin{equation}\label{eq-universalidad-10}
  \begin{split}
   \ket{n^{\infty}(t)} &= U^{\infty}(t) \ket{n^{\infty}(0)}
    \\
    &= e^{-\frac{i}{\hbar}H^{\mr{ISW}}t}\ket{n^{\infty}(0)} \enspace,
 \end{split}\end{equation}
 where $U^{\infty}(t)$ is the unitary evolution operator and $\ket{n^{\infty}(0)}=\sum_{n} c_n\ket{n}^{\infty}$. The $U^{\infty}(t)$ matrix elements are 

 \begin{equation}\label{eq-universalidad-15}
   U_{np}^{\infty}(t)=\exp \left (-i\frac{\hbar n(n+2)t}{2ma^2} \right ) \delta_{np} \enspace ,
 \end{equation}
 where $\delta_{np}$  is the Kronecker delta.

 From different choices of the $t$ parameter in \eqref{eq-universalidad-15}, and different qubits coding, it is possible to find the evolutions corresponding to different quantum gates. In order to build the $CNOT^{\infty}$ gate, based on the normalized eigenvectors of \eqref{eq-universalidad-01}, we code the basis for a $2$-qubit $\left \{ \ket{00}^4,\ket{01}^4, \frac{\ket{10}^4+\ket{11}^4 }{\sqrt{2}},\frac{\ket{11}^4-\ket{10}^4 }{\sqrt{2}} \right \}$ in the canonical basis \eqref{eq-5} by  
 \begin{equation}\label{eq-universalidad-20}
   \begin{aligned}
\ket{00}^4  & \xrightarrow{\mr{codification}} \ket{0}^{\infty}  \enspace, &  \qquad \ket{01}^4   & \xrightarrow{\mr{codification}} \ket{2}^{\infty} \enspace , 
 \\
 \frac{\ket{10}^4+\ket{11}^4 }{\sqrt{2}}  & \xrightarrow{\mr{codification}} \frac{\ket{4}^{\infty}+\ket{1}^{\infty} }{\sqrt{2}} \enspace , & \qquad \frac{\ket{11}^4-\ket{10}^4 }{\sqrt{2}} &  \xrightarrow{\mr{codification}} \frac{\ket{4}^{\infty}-\ket{1}^{\infty}
  }{\sqrt{2}} \enspace .
 \end{aligned}
\end{equation}

According to \eqref{eq-universalidad-15}, in  $t=\frac{2\pi ma^2}{\hbar}$ we obtain that $CNOT^{\infty} = U^{\infty} \left ( \frac{2mL^2}{\hbar\pi^{2}} \right )$, where the $CNOT^{\infty}$ matrix elements are $(CNOT^{\infty})_{np} = (-1)^n\delta_{np}$, and the transformation carried out upon the coded basis \eqref{eq-universalidad-20} is given by  
 \begin{equation*}
   \begin{aligned}
 \ket{0}^{\infty}  & \xrightarrow{CNOT^{\infty}} \ket{0}^{\infty} \enspace, & \qquad   \ket{2}^{\infty} & \xrightarrow{CNOT^{\infty}} \ket{2}^{\infty} \enspace, 
 \\ 
 \frac{\ket{4}^{\infty}+\ket{1}^{\infty}
  }{\sqrt{2}} & \xrightarrow{CNOT^{\infty}} \frac{\ket{4}^{\infty}-\ket{1}^{\infty} }{\sqrt{2}} \enspace, & \qquad \frac{\ket{4}^{\infty}-\ket{1}^{\infty}}{\sqrt{2}}&\xrightarrow{CNOT^{\infty}}\frac{\ket{4}^{\infty}+\ket{1}^{\infty}}{\sqrt{2}} \enspace.
 \end{aligned}
 \end{equation*}

Given that the universal minimal sets of quantum gates contain gates of $1$-qubit and the $CNOT^4$ gate, it is not possible to carry out a codification of the gates of $1$-qubit over the same basis used for the $CNOT^4$ gate, therefore, it is not possible to obtain the universality of our hypercomputation model parting from compound universal sets via gates that act upon a different number of qubits\footnote{We thank to an anonymous referee for having clarified this point.}. A possibility would be to obtain the TM-universality parting from unitary sets of universal gates, such as the Toffoli or Fredkin gates. This will be investigated elsewhere.



\section{Conclusions}
It is quite surprising that a quantum computation model over such a \emph{simple} physical system like the infinite square well has hypercomputational characteristics as it was shown. The success obtained by choosing an underlying physical system different from the one selected by Kieu opens the possibility of obtaining new hypercomputation models supported by the quantum computation. 

Kieu's algorithm does not depend on any intrinsic peculiarity of the simple harmonic oscillator nor on its dynamical algebra $\mf{g}_{\mr{W \negthickspace - \negthickspace H}}$. It is possible to consider that in principle any finite-dimensional dynamical algebra that admits an infinite-dimensional irreducible representation, and that it admits the formation of attainable coherent states in a known quantum system, is a good candidate to establish an algorithm \emph{à la} Kieu.


\acknowledgments
We thank Prof. Tien D. Kieu for helpful discussions and feedback. One of us (A. S.) would likes to acknowledgment the kind hospitaly during his visit to Prof. Kieu at the CAOUS at Swinburne University of Technology. We thank to some anonymous referees for their accurate observations and suggestions to preliminary versions of this article. This research was supported by COLCIENCIAS (grant \#RC-284-2003) and by EAFIT University (grant \#1216-05-13576).


\newcommand{\filehomepage}[1]{http://sigma.eafit.edu.co:90/~asicard/archivos/#%
1}

\end{document}